\documentclass[reprint,
 amsmath,amssymb,
 aps,superscriptaddress
]{revtex4-2}

\usepackage{graphicx}
\usepackage{dcolumn}
\usepackage{bm}
\usepackage{comment}
\usepackage[dvipsnames]{xcolor}

\begin{document}

\graphicspath{{./images}}

\preprint{APS/123-QED}

\title{Mapping the topological proximity-induced gap in multiterminal Josephson junctions}

\author{M. Wisne}
\affiliation{Department of Physics and Astronomy, Northwestern University, 2145 Sheridan Road, Evanston, IL 60208, USA}
\author{Y. Deng}
\affiliation{Department of Physics and Astronomy, Northwestern University, 2145 Sheridan Road, Evanston, IL 60208, USA}
\author{I. M. A. Lilja}
\affiliation{QTF Centre of Excellence, Department of Applied Physics, Aalto University, Espoo, Finland}
\author{P. J. Hakonen}
\affiliation{QTF Centre of Excellence, Department of Applied Physics, Aalto University, Espoo, Finland}
\author{V. Chandrasekhar}
\thanks{Corresponding author\\email: v-chandrasekhar@northwestern.edu}
\affiliation{Department of Physics and Astronomy, Northwestern University, 2145 Sheridan Road, Evanston, IL 60208, USA}

\date{\today}

\begin{abstract}
Multiterminal Josephson junctions (MTJJs), devices in which a normal metal is in contact with three or more superconducting leads, have been proposed as artificial analogs of topological crystals.  The topological nature of MTJJs manifests as a modulation of the quasiparticle density of states (DOS) in the normal metal that may be probed by tunneling measurements.  We show that one can reveal this modulation by measuring the resistance of diffusive MTJJs with normal contacts, which shows rich structure as a function of the phase differences $\{\phi_i \}$.  Our approach demonstrates a simple yet powerful technique for exploring topological effects in MTJJs.    
\end{abstract}

\maketitle
The energy of an Andreev bound state formed in a ballistic normal metal connected to two superconducting contacts disperses with the phase difference $\phi$ between the superconductors similar to how the energy of an electron in a one-dimensional crystal disperses with the crystal momentum $k$.  This similarity has led to the suggestion of creating artificial analogs of $n$ dimensional crystals with a normal metal in contact with $n+1$ superconductors with $n$ distinct phase differences $\{\phi_i \}$, with predictions that the resulting band structure may be topologically non-trivial \cite{riwar_multi-terminal_2016}. The ability to access different topological regions simply by tuning the phase differences $\{\phi_i \}$ is an attractive feature of this system. In the ballistic case, theory predicts topologically distinct gapped regions in the ``momentum'' space defined by the phases $\{\phi_i \}$ separated from each other by regions where the gap closes, as well as quantized conductance between the superconducting contacts in the gapped regime in direct analogy with real topological materials \cite{riwar_multi-terminal_2016, eriksson_topological_2017, xie_weyl_2018}. While the modulation of the gap has been observed in tunneling measurements \cite{coraiola_phase-engineering_2023, coraiola_spin-degeneracy_2023}, quantized conductance in transport measurements has not been observed to our knowledge \cite{graziano_transport_2020, arnault_multiterminal_2021, arnault_dynamical_2022}. 

If the normal metal is diffusive, one does not expect well-defined Andreev levels. In this case, the topological behavior is tied to the winding numbers of the phase of the quasiclassical Green's function describing the proximitized normal metal \cite{strambini_-squipt_2016, amundsen_analytically_2017, vischi_coherent_2017}.  A modulation of the quasiparticle DOS with the phases $\{\phi_i\}$ is also predicted, with different gapped regions in the quasiparticle DOS defined by unique topological indices separated by regions where the gap closes. Tunneling measurements on a diffusive device with three superconducting contacts have indeed shown the predicted modulation of the DOS with the phases $\phi_1$ and $\phi_2$, albeit only along the line $\phi_2=-\phi_1$ as the phase modulation was achieved using a uniform external magnetic field \cite{strambini_-squipt_2016}.

\begin{figure*}
\includegraphics[width = 18 cm]{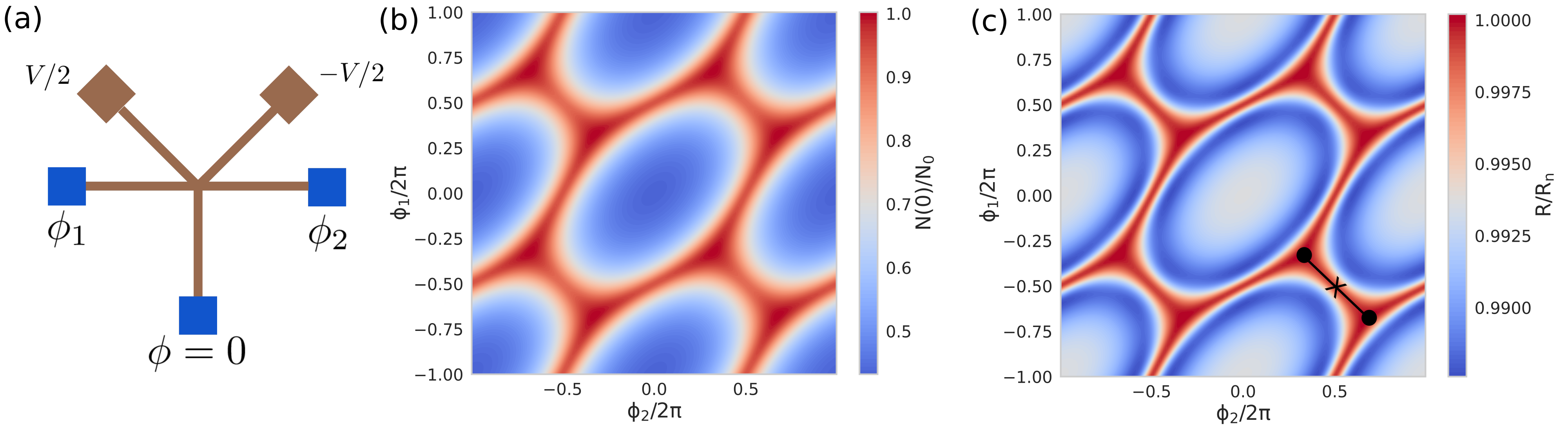}
\caption{\label{fig:fig1} (a) Schematic of the sample used for the numerical simulations.  Gold represents the normal metal, with the squares being the normal contacts, while blue represents the superconducting contacts.  A voltage $V$ is applied symmetrically between the two normal contacts. The phases of the superconducting contacts are specified as shown. (b) Density of states at the Fermi energy $N(0)$ at the junction of all the normal wires normalized to the normal state value $N_0$ as a function of the phase differences $\phi_1$ and $\phi_2$  (c) Low temperature resistance $R$ between the two normal contacts normalized to the normal state resistance $R_n$.}
\end{figure*}

In our experience, fabricating devices for tunneling measurements is challenging; it is easier to fabricate devices for electrical transport measurements.  The initial theoretical studies for diffusive systems were for a normal metal connected only to superconducting contacts where the modulation of the DOS is uniform throughout the normal metal.  To perform electrical transport measurements on the proximity-coupled normal metal itself, one also needs to attach normal contacts.  In this case, the DOS is no longer uniform in the normal metal but varies as a function of position between the normal contacts.  Nevertheless, with appropriately designed devices, there can be a region in the diffusive normal metal where the DOS mimics the DOS of a device with only superconducting contacts.  More interestingly, measurements of the resistance of the \emph{entire} device reflect this DOS \cite{chandrasekhar_probing_2022}. This is demonstrated in Fig. \ref{fig:fig1}, which shows the DOS and resistance of a diffusive normal metal with three superconducting contacts and two normal contacts (Fig. \ref{fig:fig1}(a)) as a function of the phase differences $\phi_1$ and $\phi_2$ between the superconducting contacts, calculated using the quasiclassical superconducting Green's function technique. The parameters used in this simulation are based on Sample 1 of this study:  details of the calculations can be found in the \textit{End Matter} and in Ref. \citenum{chandrasekhar_probing_2022}.  The DOS at the juncture of the normal metal wires (Fig. \ref{fig:fig1}(b)) is very similar to the earlier predictions for diffusive normal metals connected to only 3 superconducting contacts \cite{strambini_-squipt_2016, amundsen_analytically_2017}, with a large modulation of the DOS and gap closings at specific points in the phase space.    Calculations of the resistance between the two normal reservoirs (Fig. \ref{fig:fig1}(c)) closely track this DOS, with the resistance achieving its normal state value $R_n$ when the gap in the DOS closes, although there is more structure in the resistance in comparison to the DOS.

\begin{figure*}
     \includegraphics[width=18cm]{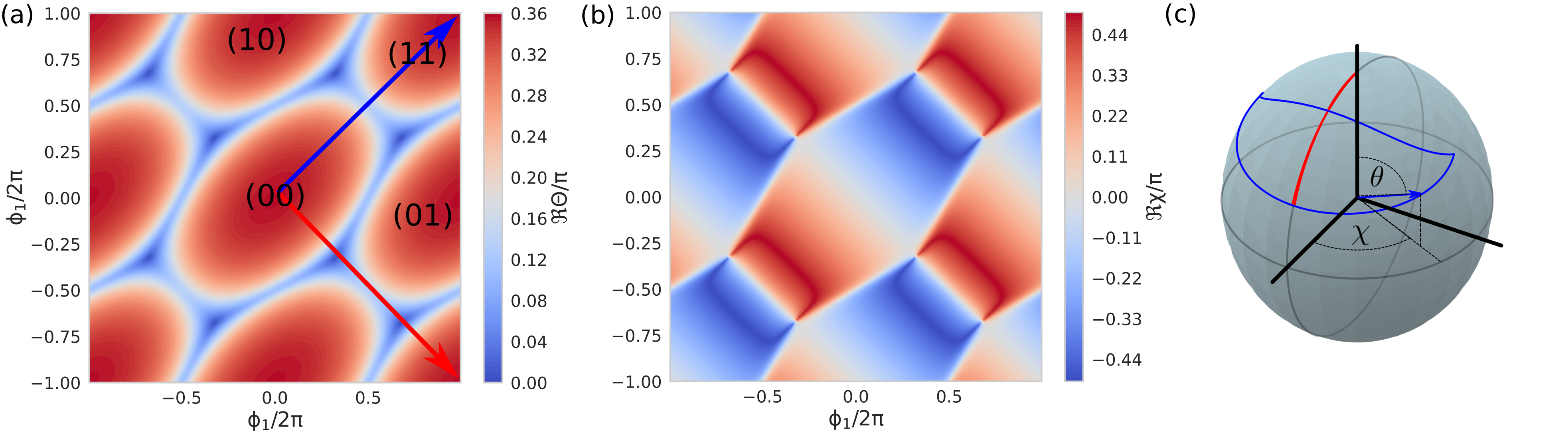}
     \caption{ (a) (b)  Real part of $\theta$ (a) and $\chi$ (b) as the junction of all the normal wires (see Fig. \ref{fig:fig1}(a)) as a function of $\phi_1$ and $\phi_2$.  $(n_1n_2)$ in (a) identify the topological indices that characterize different gapped regions.  (c)  Mapping of the blue and red trajectories shown in (a) on the Bloch sphere where the polar angle represents $\Re \; \theta$ and the azimuthal angle represents $\Re \; \chi$, and where the north pole denotes the completely normal state.}
     \label{fig:fig2}
\end{figure*}

The modulation of the DOS and resistance with $\{\phi_i \}$ reflects the changes in the retarded Green's function $\hat{G}^R$ that describes the proximitized normal metal.   $\hat{G}^R$ can be parameterized by two complex parameters $\theta$ and $\chi$ \cite{BELZIG19991251}  
\begin{equation}
\hat{G}^R= 
\begin{pmatrix}
    \cos \theta & \sin \theta \; e^{i \chi} \\
    \sin \theta \; e^{- i \chi} & -\cos \theta
\end{pmatrix}.
\label{eqn:eqn1}
\end{equation}
Here, $\theta$ and $\chi$ characterize the strength of the superconducting pair correlations and their gauge-invariant phase, respectively.  $\theta = \pi/2$ at a superconducting contact and vanishes at a normal contact, while $\chi$ is set at the superconducting contacts by the imposed phases $\{\phi_i \}$.  In the proximitized normal metal, $\theta$ and $\chi$ will vary between these values.  As pointed out by Strambini \textit{et al.} \cite{strambini_-squipt_2016}, the resulting state in the proximitized normal metal can be mapped on the northern hemisphere of a Bloch sphere, with $\theta$ representing the polar and $\chi$ representing the azimuthal angle.  The equator of the Bloch sphere represents the fully gapped (superconducting) state while the North pole represents the gapless (normal) state.  As our device has normal contacts in addition to superconducting contacts, $\theta$ will approach but never equal $\pi/2$ in the proximitized normal metal.  Nevertheless, the overall behavior of $\theta$ and $\chi$ is similar to that expected in a diffusive metal with only superconducting contacts \cite{strambini_-squipt_2016, amundsen_analytically_2017}. This is demonstrated in Fig. \ref{fig:fig2}(a) and (b), which shows the real parts of $\theta$ and $\chi$ in the center of the wire of Fig. \ref{fig:fig1}(a) as a function of $\phi_1$ and $\phi_2$ (the imaginary parts are effectively zero to numerical accuracy).  There are regions with large $\theta$ corresponding to the gapped areas in Fig. \ref{fig:fig1}(b). We label these regions with topological indices $(n_1 n_2)$ corresponding to the winding numbers of the phases $\phi_1$ and $\phi_2$ \cite{strambini_-squipt_2016}, as shown in Fig. \ref{fig:fig2}(a).  These regions are separated by narrow sections where $\theta \sim 0$ and $\chi=0$, corresponding to the areas of phase space where the system approaches the normal state.  One can map trajectories between adjacent minima in this phase space on the Bloch sphere shown in Fig. \ref{fig:fig2}(c).  The blue trajectory in Fig. \ref{fig:fig2}(c) corresponds to the blue trace in Fig. \ref{fig:fig2}(a) from the minimum specified by the winding numbers (00) to the minimum specified by the winding numbers (11), while the red trajectories correspond to going from (00) to (-11).  In the case of the blue trajectory, both $\theta$ and $\chi$ vary, while for the red trajectory $\chi$ is always 0, and hence the red trajectory on the Bloch sphere appears an arc of a meridian.  Note that for both trajectories, the maximum value of $\theta$ is $\theta \sim 0.36 \;\pi$ rather than $0.5 \; \pi$ as would be expected for a MTJJ with no normal contacts.  In addition, the blue trajectory does not approach the north pole of the Bloch sphere corresponding to the completely normal state, while the red trajectory does (twice, in fact, although this is not discernible as $\chi$ is always 0). $\theta$ can be pushed closer towards $\pi/2$ by decreasing the distance between the superconducting contacts and the junction of all the normal wires, but it will never equal $\pi/2$ due to the presence of the normal contacts.  However, the main point is that both trajectories shift from the maximum value of $\theta \sim 0.36 \; \pi$ corresponding to the gapped state in going between minima identified by different winding numbers.      

This analysis shows that resistance measurements on proximity-coupled devices are an attractive route to explore topological effects in diffusive MTJJs.  To this end, we fabricated and measured three devices based on the geometry of Fig. \ref{fig:fig1}(a) for this study:  we discuss the results for two (Sample 1 and Sample 2) for which we have the most complete sets of data.   These two devices differed slightly in sample dimensions as well as in the thickness of the normal metal (see the Supplementary Materials for details).  Figure \ref{fig:fig3}(a) shows a SEM image of Sample 1.  In order to independently and statically vary the phase between the superconducting contacts and thus map out the entire phase space $\{\phi_1, \phi_2\}$, the superconducting contacts were connected to form two loops through which a magnetic field could be threaded by on-chip field coils generating a flux, as shown schematically in Fig. \ref{fig:fig3}(a).  The area of each flux loop was $\sim 5$ $\mu$m $\times$ 50 $\mu$m.  Measurements on a device in which one flux loop was broken showed that each field coil coupled only to its respective flux loop, with no measurable flux coupled to the other loop within the range of field coil drive currents used.  The resistance of the device was measured with standard 4-terminal low-frequency ac lock-in techniques using the contacts labeled on the left of Fig. \ref{fig:fig3}(a).  Devices included in this study were fabricated using e-beam lithography on Si substrates with a 1 $\mu$m SiO$_2$ insulating layer. All features were patterned in MMA/PMMA bilayers using a Tescan MIRA 4 electron microscope onto which 99.999\% (5N) Au and Al films were deposited in an Edwards thermal evaporator used exclusively for ultra pure Au and Al. Prior to the Au deposition, an \textit{in situ} O$_2^+$ plasma etch was used to clean the substrate. Additionally, to ensure high NS interface transparency, an \textit{in situ} Ar$^+$ plasma etch was performed immediately prior to a deposition of Al, leaving only enough time to pump back to a base pressure of $8 \times 10^{-7}$ torr. The devices were cooled in an MX100 Oxford dilution refrigerator within hours of the Al deposition to prevent degradation of the NS interfaces. 

Measurements were performed using PAR 124 lock-in amplifiers with a modified Adler-Jackson resistance bridge using low-frequency ($<$ 120 Hz) ac excitation currents of amplitudes of $\sim 1$ $\mu$A.  The ac excitation current was chosen experimentally to optimize the signal to noise without modifying the measured response.  Field coil bias was provided by two home-built current sources driven by separate Agilent synthesizers.   To avoid interference from line-frequency noise, the battery-operated first-stage pre-amplifiers were placed in a $\mu$-metal shielded enclosure attached to the cryostat.

Two length scales in the normal metal determine the overall dimensions of the devices: the electron phase coherence length $L_\phi$ and the superconducting coherence length $\xi_N = \sqrt{\hbar D/k_B T}$ \cite{BELZIG19991251}.  Here $T$ is the temperature and $D=(1/3) v_F \ell$ the electronic diffusion coefficient, $v_F$ being the Fermi velocity and $\ell$ the elastic scattering length of the electrons in the normal metal.  Fits to the weak localization magnetoresistance (MR) \cite{BERGMANN19841} of co-deposited Au meander wires gave $L_\phi$ in excess of 7 $\mu$m at base temperature ($\sim$ 20 mK), so that the devices were entirely phase coherent in the measurement temperature range.  Sheet resistance measurements of the devices gave $\xi_N \sim 0.5 \mu$m$/\sqrt{T}$, with $T$ in Kelvin.  Thus $\xi_N \sim 3$ $\mu$m at base temperature, also larger than the sample dimensions.

\begin{figure*}
     \includegraphics[width=18cm]{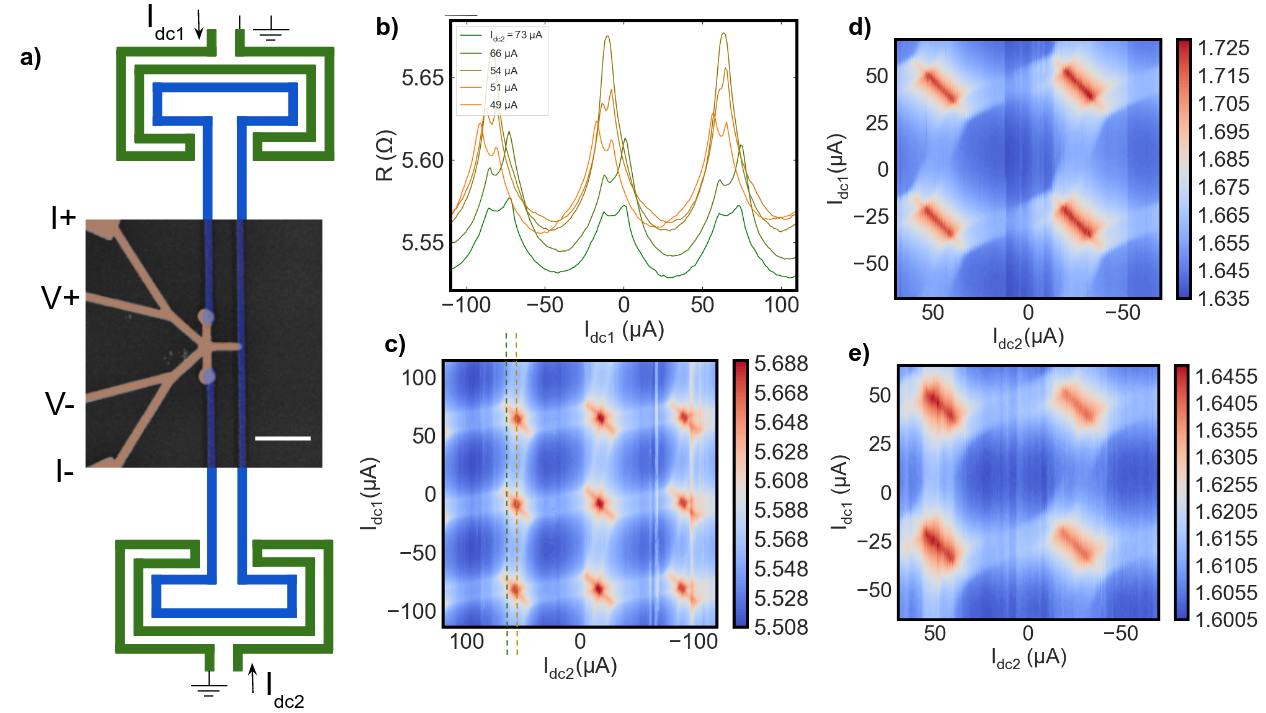}
     \caption{ (a) False-color SEM image of Sample 1. Gold (blue) corresponds to Au (Al). The scale bar is $1 \ \mu$m. The schematic shows the superconducting (Al) flux loops attached to the superconducting contacts, as well as the on-chip Al field-coils (green). The leads used for the 4-terminal resistance measurement are labeled. (b)  Resistance of Sample 1 as a function of coil bias $I_{dc1}$ at a few different values of $I_{dc2}$ at 600 mK. (c)  Contour map of the resistance in ohms of Sample 1 at 600 mK as a function of $i_{dc1}$ and $I_{dc2}$.  Vertical dotted lines show cuts corresponding to $I_{dc2} = 73$ $\mu$A  and $I_{dc2}=54$ $\mu$A in (b).  (d), (e) Contour map of the resistance of Sample 2 at 300 mK and 30 mK, respectively.   
     \label{fig:fig3}} 
\end{figure*}

Ideally, one would like the length of the normal metal wires between the superconducting contacts and the node at which these wires meet to be as short as possible in order to maximize the proximity effect and push the value of $\theta$ close to $\pi/2$, corresponding to the fully gapped state as discussed above.  However, this leads to hysteresis in the response as a function of the flux threaded through each flux loop, as the large critical current of the proximitized region coupled with the self-inductance of the flux loops results in screening of the externally applied flux.  In the first device we measured (Sample 1), the length between superconducting contacts $L$ was $\sim1$ $\mu$m, and the resistance traces of this device at base temperature were indeed hysteretic (see Supplementary Materials).  The critical current of the proximitized region depends exponentially on the ratio $L/\xi_N$ \cite{Dubos_PhysRevB.63.064502}.  Since $\xi_N$ decreases with increasing temperature, we can reach a non-hysteretic regime by raising the temperature.  The lowest temperature at which this should occur is when $L\sim \xi_N$.  Although we cannot directly measure the critical current between the superconducting contacts, from a comparison of the period of the oscillations seen at base temperature to the range over which no oscillations are observed on switching the sweep direction of the field coil, one can roughly estimate the temperature at which $L\sim \xi_N$.  From the data shown in the Supplementary, we estimated this temperature to be $\sim$600 mK for Sample 1.

Figure \ref{fig:fig3}(b) shows the resistance of Sample 1 as a function of the current $I_{dc1}$ through one field coil at a few different values over a narrow range of the current $I_{dc2}$ through the other field coil at a temperature of 600 mK.  Clear periodic oscillations as a function of $I_{dc1}$ are observed that evolve systematically with $I_{dc2}$, with no hysteresis.  The oscillations in general show an asymmetric, double-peaked structure, with the asymmetry evolving as $I_{dc2}$ is varied over a narrow range.  The peaks appear to merge into a single maximum for $I_{dc2}=54$ $\mu$A.  To obtain the full phase diagram, we swept $I_{dc1}$ over the range shown in Fig. \ref{fig:fig3}(b) while stepping $I_{dc2}$ sequentially to obtain the contour map shown in Fig. \ref{fig:fig3}(c). Small current offsets likely arise from the remanent field of the superconducting solenoid of the dilution refrigerator.  The contour map is similar overall to the results of the simulation shown in Fig. \ref{fig:fig1}(c), with structure that is periodic in both $I_{dc1}$ and $I_{dc2}$. Each period corresponds to one superconducting flux quantum $\Phi_0 = h/2e$ through an individual flux loop.  As with the simulations, there are central areas of low resistance separated by regions of higher resistance.  In particular, there are localized, periodic regions of high resistance; these correspond to the diagonal line shown in Fig. \ref{fig:fig1}(c), and are responsible for the evolution of the traces with $I_{dc2}$ seen in Fig. \ref{fig:fig3}(b).  While the overall structure is similar, there are also significant differences.  First, while the unit cell of the structure in the simulations of Fig. \ref{fig:fig1}(c) is roughly elliptical in shape, it is rhomboidal in the experimental data seen in Fig. \ref{fig:fig3}(c).  Second, a key feature of the simulations is that the maxima in resistance (and correspondingly the maxima in $N(0)$) occur \textit{not} at the points defined by $\phi_1,\phi_2 = (2n + 1) \pi$ (marked with the cross on the diagonal line in Fig. \ref{fig:fig1}(c)), but at pairs of points in phase space a short distance away along a diagonal (marked with filled circles in Fig. \ref{fig:fig1}(c)).  In the experimental data, these diagonals of high resistance can be observed, but the maxima in resistance do appear to occur at field coil currents corresponding to $\phi_1,\phi_2 = (2n + 1) \pi$.  Finally, in the simulations, the regions of high resistance are connected to each other by narrow strips in phase space where the resistance almost reaches its maximum value.  The experimental data also shows similar strips, but the resistance does not approach the maximum resistance seen near $\phi_1,\phi_2 = (2n + 1) \pi$.  These regions correspond to the (almost) symmetrical double peaked oscillations shown in Fig. \ref{fig:fig3}(b) for $I_{dc2}=49$ $\mu$A.

\begin{figure*}
\centering
     \includegraphics[width=14cm]{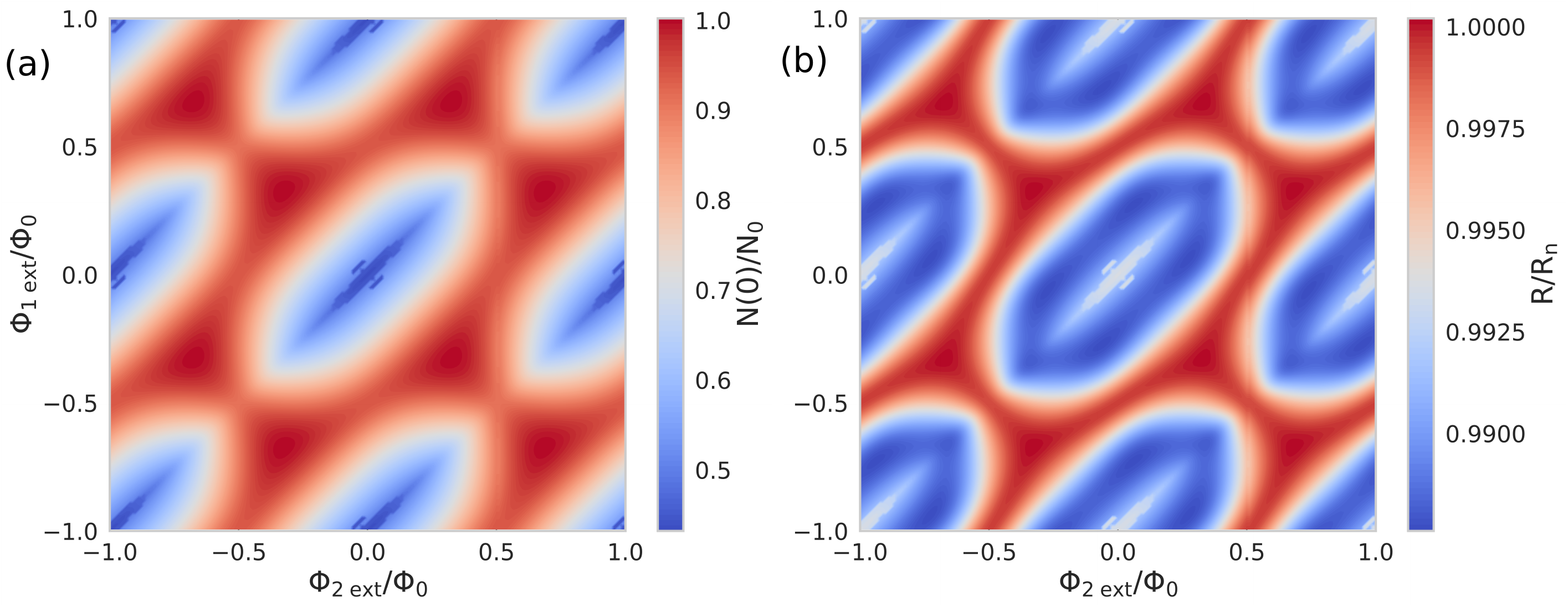}
     \caption{(a), (b) Normalized DOS and resistance, respectively, for the geometry of Fig. \ref{fig:fig1}(a) determined by calculating the phase on the superconducting contacts self-consistently using Eqn. (\ref{eq:eq2}).}  
     \label{fig:fig4}
\end{figure*}
To confirm the reproducibility of these results, we fabricated and measured a second set of devices of the same design. In order to make measurements at the base temperature of our refrigerator without encountering hysteresis as noted above, these devices were fabricated with a length $L\sim 2$ $\mu$m between the superconducting contacts.  Figures \ref{fig:fig3}(d) and (e) show the phase diagrams for one of these devices (Sample 2) at 300 mK and 30 mK respectively.  These data are similar to those of Sample 1 shown in Fig. \ref{fig:fig3}(c), except that the distinct maxima seen at $\phi_1,\phi_2 = (2n + 1) \pi$ are no longer visible.  Sample 2 shows only diagonal lines of high resistance, more consistent with the simulations.  There is no significant difference in the structure of the phase diagram between 30 mK and 300 mK, although the overall resistance is higher as expected, and the modulation of the resistance is larger at higher temperatures (5.2\% compared to 2.7\%).  This is consistent with the numerical simulations (see Supplementary Materials).

The results of the simulations shown in Fig. \ref{fig:fig1} are obtained by imposing the phases $\{\phi_1, \phi_2\}$ at the superconducting contacts and the voltage at normal contacts before calculating the currents through all normal metal wires.  In general, the phases $\phi_i$ need to be determined self-consistently by applying the usual flux quantization condition for each flux loop \cite{tinkham2004introduction}

\begin{equation}
    \phi_i + 2 \pi \frac{\Phi_i}{\Phi_0} = \oint \nabla \phi \cdot d \mathbf{l} = 2\pi n_i
    \label{eq:eq2}
\end{equation}
where $n_i$ is the topological index introduced by Strambini \textit{et al.} \cite{strambini_-squipt_2016}, $\phi_i$ is the (gauge-invariant) phase difference between the superconducting contact and the reference contact ($\phi=0$) and $\Phi_i$ is the \textit{net} flux enclosed by the corresponding loop.  If the self-inductance $L_i$ of the flux loop is not negligible, $\Phi_i$ is not equal to the flux $\Phi_{ext}$ generated by the on-chip field coils, but is given by $\Phi_i = \Phi_{ext} - L_i I_{si}$, $I_{si}$ being the supercurrent flowing through the flux loop.  In the experiments, we estimate the self-inductance of each flux loop to be approximately 0.1 nH and the screening supercurrents to be on the order of $\mu$A based on prior experience with similar proximity coupled devices \cite{Noh_PhysRevB.88.024502}, so that the flux generated can be an appreciable fraction of $\Phi_0$.  We have clearly seen this screening effect in Sample 1 at low temperatures (see Supplementary Materials). In order to take the effects of the screening current into account, we have simulated the DOS and resistance of the device of Fig. \ref{fig:fig1}(a) by self-consistently calculating the phases $\phi_1$ and $\phi_2$ on each superconducting contact using Eqn. (\ref{eq:eq2}).  Theoretically, the maximum value of the critical supercurrent in a SNS junction in the long junction limit is $I_c \sim 10.82 E_c/eR_n$ assuming perfectly transparent normal-metal/superconductor interfaces \cite{Dubos_PhysRevB.63.064502}. Here $E_c = \hbar D /L^2$ is the Thouless energy, $L$ being the length of the normal metal between the two superconductors.  In previous experiments, we have found that the critical current is typically a factor of $\sim20$ smaller than this, likely due to the fact that the interface transparency is finite \cite{Noh_PhysRevB.88.024502}. Since the numerical simulations, which assume perfect normal-metal/superconductor interfaces, overestimate the supercurrents in our device, we use a correspondingly smaller value of the self-inductances $L_i$ in our simulation.  The results of these simulations are shown in Fig. \ref{fig:fig4}, where now the $x$ and $y$ axes represent the normalized external flux coupled to each flux loop.  The overall structure is similar to that seen in Fig. \ref{fig:fig1}, with an elliptical region in the center of each cell where the DOS is at a minimum with correspondingly smaller resistance, but the detailed structure is modified.  In particular, the transitions between these minima are broader, in line with our experimental observations.  While a more accurate agreement between theory and experiment would likely require taking into account the inevitable asymmetries in the flux loops as well as the different transparencies of each normal-superconducting interface, it is clear that it is important to take into account the self-inductance of the flux loops in such devices.   Regardless, we note that the overall topological structure is similar to that of Fig. \ref{fig:fig1} and is thus robust against such perturbations.

In conclusion, we have measured the resistance of diffusive three-terminal Josephson junctions as a function of the two independent superconducting phase differences imposed by on-chip field coils.  The resistance shows a rich structure as a function of these phase differences, reflective of the topological nature of the underlying anomalous superconducting Green's function. The overall structure of the phase diagram is consistent with numerical predictions based on the quasiclassical theory of superconductivity, although the detailed structure is different, likely due to screening effects arising from the finite self-inductance of the flux loops used to generate the superconducting phases.  Our results demonstrate the power of using electrical transport measurements with normal contacts to explore the physics of MTJJs, including potentially in ballistic devices where a different class of topological effects is predicted. 

\textit{Acknowledgements}--This research was conducted with support from the National Science Foundation under grant No. DMR-2303536 and from the European Union’s Horizon 2020 Research and Innovation Programme, under Grant No. 824109 (EMP). IMAL and PJH acknowledge support by the Research Council of Finland Project Nos. 341913 (EFT), 352926 (QTF), and by the Jane and Aatos Erkko Foundation and the Keele Foundation (SuperC project). \\

\noindent\textbf{End Matter}\\
\noindent\textit{Numerical Simulations}

The simulations for this paper were performed by solving the quasiclassical equations of superconductivity in the Keldysh formulation, which allows one to obtain solutions for the superconducting Green's functions (the Usadel equation) as well as for the quasiparticle distribution functions (the kinetic equations) \cite{BELZIG19991251}. The equations are conveniently solved using the software package developed by Pauli Virtanen \cite{usadel1}. Details of these numerical calculations can be found in Ref. [\citenum{chandrasekhar_probing_2022}].  In brief, for the simulations shown in Figs. \ref{fig:fig1} and \ref{fig:fig2}, referring to the schematic in Fig. \ref{fig:fig1}(a), we set the phase of one superconducting contact to 0 without loss of generality, and assign phases $\phi_1$ and $\phi_2$ to the remaining two superconducting contacts. A small voltage bias $V$ is applied symmetrically between the two normal contacts.  The normal-superconducting contacts were assumed to be perfectly transparent.  Solution of the Usadel equation enables one to determine the normalized density of states at any point along the normal wire, and in particular at the junction of all the normal wires, which is what is plotted in Fig. \ref{fig:fig1}(b).  Solution of the kinetic equations allows one to calculate the quasiparticle current flowing into the normal contact and from this the normalized resistance between the normal contacts plotted in Fig. \ref{fig:fig1}(c).  The parameters used in the simulations are based on the dimensions and film parameters of Sample 1.  In particular, the Thouless energy $E_c= \hbar D/L^2$ which sets the energy scale of the proximity effect is determined by the length $L$ of the normal wire between the voltage probes (see Fig. \ref{fig:fig3}(a)).

For the self-consistent calculation whose results are shown in Fig. \ref{fig:fig4}, the Usadel ad kinetic equations were solved iteratively until the condition given by Eqn. (\ref{eq:eq2}) was satisfied.  

\bibliography{dATP}

\end{document}

\title{Supplementary Materials: Mapping the topological proximity-induced gap in multiterminal Josephson junctions}

\author{M. Wisne,$^1$ Y. Deng,$^1$ I. M. A. Lilja,$^2$ P. J. Hakonen,$^2$ and V. Chandrasekhar$^1$}

\footnotetext[1]{Physics and Astronomy Department, Northwestern
University, Evanston, IL 60208, USA}
\footnotetext[2]{QTF Centre of Excellence, Department of Applied Physics, Aalto University, Espoo, Finland}

\baselineskip24pt

\maketitle

\section{Estimation of $L_\phi$ and $\xi_N$}

\begin{figure}
     \begin{center}
     \includegraphics[width=10cm]{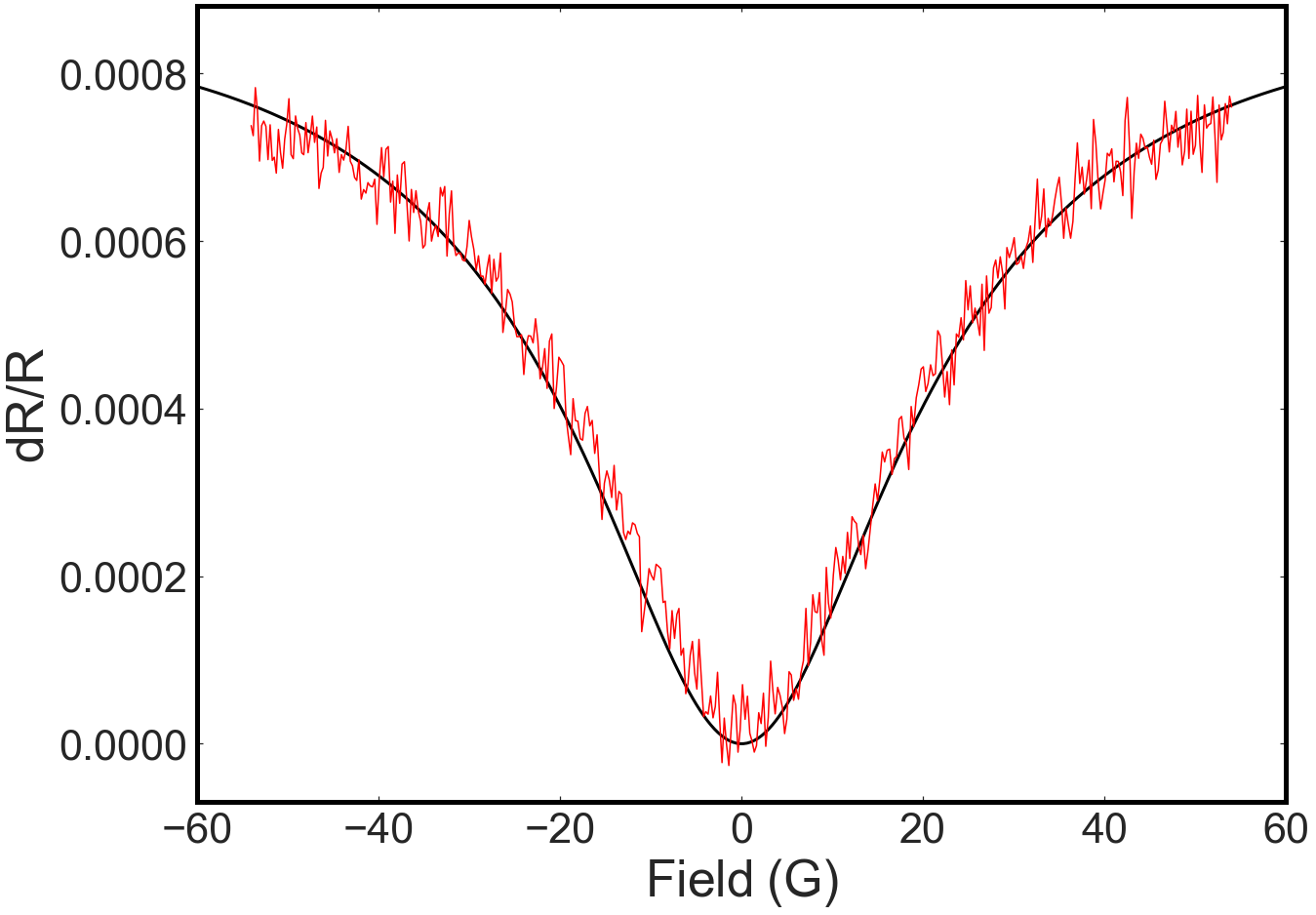}
     \caption{\textbf{Magnetoresistance weak antilocalization fitting.} The measured 4-terminal resistance (red) of a 85 nm wide, 220 $\mu$m long wire co-evaporated with Sample 1 is normalized to its value at zero field where the effect of weak localization is maximal. Weak antilocalization fitting (black) yields $L_{\phi} \sim 7.5 \ \mu$m.   }
     \label{fig:MR}
    \end{center}
\end{figure}

There are two length scales that constrain the geometry of the device: the normal electron phase coherence length $L_\phi$ and the superconducting coherence length (or Thouless length) $\xi_N$ characterizing the diffusion of pair correlations in proximity-coupled gold. Both of these lengths should be greater than the distance $L$ between superconducting contacts for coherent effects to coalesce into a measurable change in the DOS. $L_{\phi}$ and $\xi_N$ were determined by measuring the low-temperature properties of a simultaneously deposited long wire of gold of width and thickness equal to the normal metal sections of the actual device. $L_\phi$ is determined by fitting the low-field magnetoresistance (MR) of the gold wire to the theory of weak localization with spin-orbit interactions.\cite{BERGMANN19841}  Figure \ref{fig:MR} shows an example of such a fit giving a value of $L_\phi \sim 7.5$ $\mu$m, similar to values obtained for all the measured gold wires.  From the measured low-temperature resistance as well as the measured dimensions of the long gold wire, one can obtain the sheet resistance $R_\square$, and from $R_\square$ the elastic mean free path $\ell$ and diffusion constant $D$.  For Sample 1, which had a film thickness determined by atomic force microscopy of 33 nm, we obtain $R_\square = 0.34 \ \Omega$, $\ell = 62$ nm and $D = 289 \ \text{cm}^2/\text{s}$, giving  $\xi_N = \sqrt{\hbar D/k_b T} = 0.47/\sqrt{T} \ \mu \text{m}$ with $T$ in Kelvin. Sample 2 had a film thickness of 50 nm, $R_\square = 0.2 \ \Omega$, $\ell = 85$ nm, and $D = 392 \ \text{cm}^2/\text{s}$, giving $\xi_N = \sqrt{\hbar D/k_b T} = 0.55/\sqrt{T} \ \mu \text{m}$ and a similar value for $L_\phi$.

\section{Hysteresis in the flux dependence}

\begin{figure}
     \begin{center}
     \includegraphics[width=16cm]{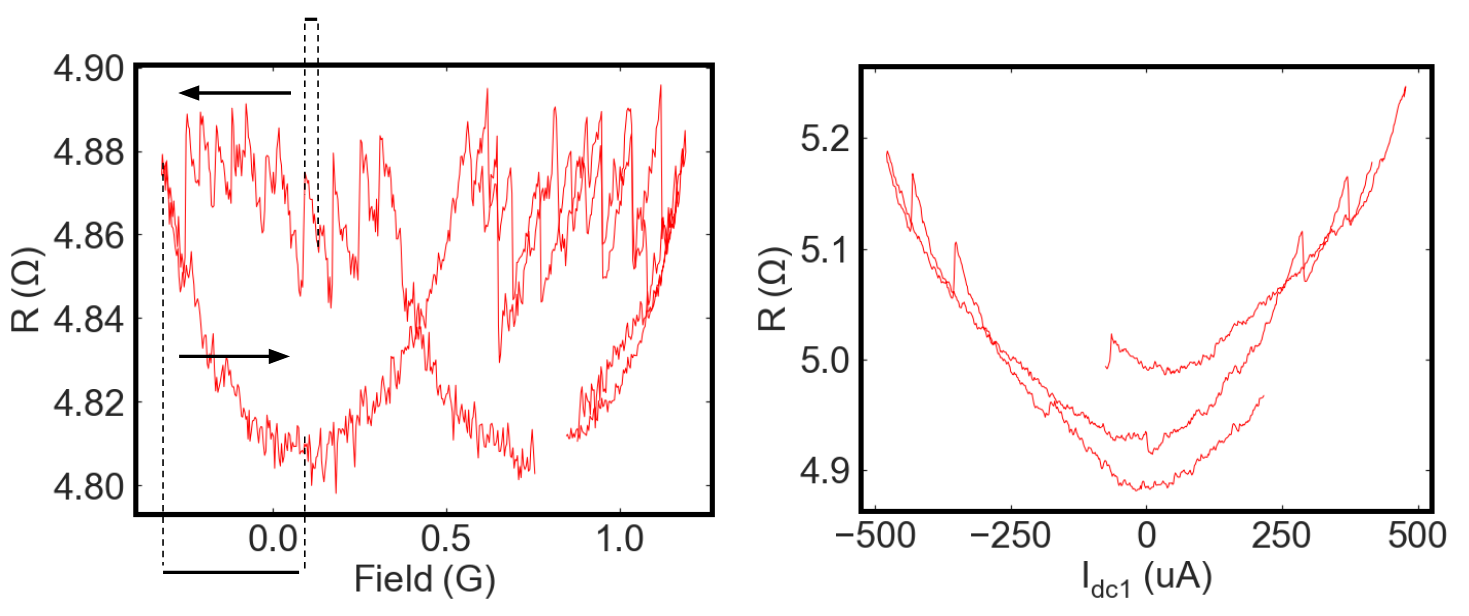}
     \caption{\textbf{Magnetoresistance (MR) hysteresis of Sample 1 at 20 mK with an external field (left) and local field (right)} Segments of MR showing half periods of oscillations for completely screened external flux and flux ``popping'' near $I_c$ are marked. Field sweep directions is shown with arrows. Flux ``popping'' is also observed with a locally applied field as abrupt changes in resistance at base temperature (right). 
     \label{fig:hys}} 
    \end{center}
\end{figure}

As noted in the main text, the response of Sample 1 was hysteretic at the base temperature of the dilution refrigerator (20 mK).  This is demonstrated in Fig. \ref{fig:hys}, which shows the resistance of Sample 1 at 20 mK with an external field (left) and field with one of the on-chip field coils (right). In both cases the response is hysteretic, reminiscent of the response of rf SQUID or superconducting loop with significant inductance.  If the field is swept to one side and then reversed, there is a broad parabolic response with no oscillations until a threshold is reached, after which one sees quasiperiodic sawtooth oscillations with period corresponding to one flux quantum $h/2e$ through the loop.  (The quasiperiodicity is due to the stochastic nature of flux jumps in the device.)  On reversing the sweep direction, a mirror symmetric response is observed.  The broad parabolic response with no oscillations is due to the essentially complete screening of the external flux by the circulating supercurrents in the flux loops, and the oscillations correspond to the entry of a single flux quantum after the critical current of the loop is exceeded.  Thus, the field or current range over which one sees the broad parabolic behavior is a measure of the critical current between the superconducting contacts, while the field or current scale of each oscillation is a measure of the supercurrent corresponding to one flux quantum.  From the left plot, from the ratio of the period of the oscillations to the range in current over which the broad parabolic response is observed, one can estimate the temperature at which the critical current $I_c$ is not sufficient to completely screen out the external field assuming an exponential temperature dependence in the long junction limit as found by Dubos \textit{et al.}\cite{Dubos_PhysRevB.63.064502}   

\section{Numerical simulations at higher temperatures}
\begin{figure}
     \begin{center}
     \includegraphics[width=16cm]{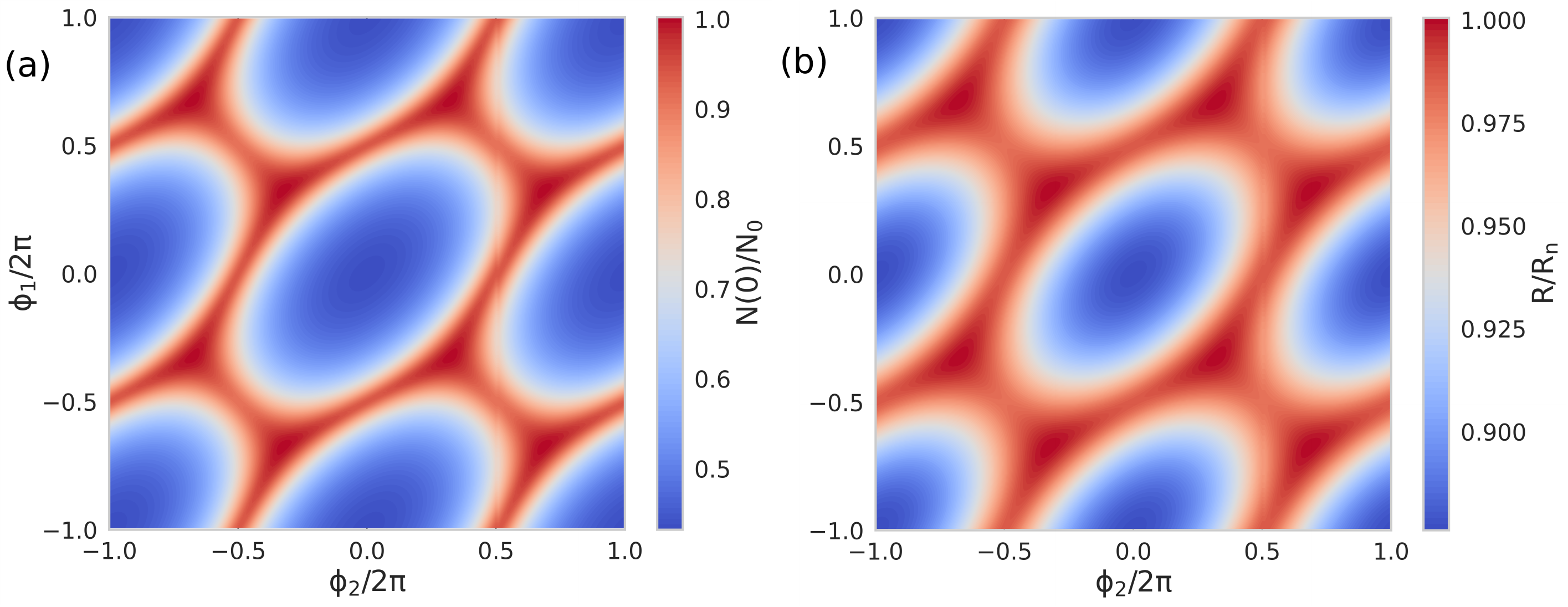}
     \caption{\textbf{Numerical simulations at higher temperatures.} Numerical calculations of the normalized DOS (\textbf{a}) and normalized resistance (\textbf{b}) for the geometry of Fig. 1(a) of the main text.  The parameters are the same as for the simulations shown in Fig. 1 of the main text, except that the temperature is now set to be equivalent to 600 mK. }  
     \label{fig:HigherTemp}
    \end{center}
\end{figure}

Figure \ref{fig:HigherTemp} shows a numerical simulation of the normalized DOS and the resistance at a temperature equivalent to 600 mK, with all other parameters being the same as for the simulation in Fig. 1 of the main text.  The DOS does not change, while the features in the resistance are less sharp as expected, but the modulation in the resistance is also higher in comparison to Fig. 1(c) of the main text, consistent with the experimental results shown in Figs. 3(d) and (e) of the main text.

\bibliographystyle{unsrt}
\normalsize
\bibliography{dATP} % Produces the bibliography via BibTeX.